\documentclass[epj]{svjour}

\usepackage{graphicx}
\usepackage{fancyhdr}
\def\makeheadbox{{
 }}
\def\mytitle{My title} 
\def\myauthors{My name}  
\def\mytype{My type of session}
\def\mysession{My session}

\def\mytitle{Physics of a pseudo-goldstone higgs from 5D} 
\def\myauthors{Adam Falkowski}    
\def\mytype{Contributed Talk}    
\def\mysession{Alternatives}

\newcommand{\cref}[1]{Chapter \ref{c.#1}}

\def\nn{\nonumber \\}  
\newcommand{\nl}{& \nonumber \\ &}

\def\beq{\begin{equation}} 
\def\eeq{\end{equation}} 
\newcommand{\ba}{\begin{array}}  
\newcommand{\ea}{\end{array}} 
\newcommand{\bea}{\begin{eqnarray}}  
\newcommand{\eea}{\end{eqnarray} }  
\newcommand{\bal}{\begin{align}}
\newcommand{\eal}{\end{align}}   
\def\bi{\begin{itemize}}  
\def\ei{\end{itemize}}  
\def\ben{\begin{enumerate}}  
\def\een{\end{enumerate}}  
\def\beq{\begin{equation}}  
\def\eeq{\end{equation}}  
\def\bc{\begin{center}}
\def\ec{\end{center}} 
 \def\bt{\begin{table}}
\def\et{\end{table}}  
 \def\btb{\begin{tabular}}
\def\etb{\end{tabular}}

\def\cl{{\mathcal L}}

\def\co{{\mathcal O}}

\def\mkk{\, M_{\rm KK}}

\def\mass2{mass${}^2$}

\def\ads{AdS$_5$\,}

\def\ra{\rangle}
\def\la{\langle}  

\def\pa{\partial}

\newcommand{\ha}{{\hat a}}

\newcommand{\ti}{\tilde}  
\def\hc{{\rm h.c.}}

\def\ov{\overline}

\begin{document}
\title{Pseudo-goldstone higgs from five dimensions}
\author{Adam Falkowski\inst{1}
}                     
%
%
\institute{
CERN, Theory Division, CH 1211, Geneva 23, Switzerland
\and 
Institute of Theoretical Physics, Warsaw University, Ho\.za 69, 00-681 Warsaw, Poland
}
%
\date{}
\abstract{
I discuss radiative generation of the higgs potential in  5D models of gauge-higgs unification.  
} 
\maketitle
%

\section{Introduction}
\label{intro}

The LHC will soon probe the mechanism that breaks electroweak symmetry and gives masses to the standard model particles.
Within the standard model, the electroweak breaking sector consists of one scalar multiplet with a potential that induces its  vacuum expectation value (vev). 
Although this simple picture is consistent with all experimental data so far, our theoretical prejudice prompts studies of more elaborate mechanisms.
The  reason is that, in the standard model, no symmetry forbids the higgs mass parameter so that it receives UV sensitive quantum corrections. Once the standard model is embedded in some more fundamental theory, it is difficult to understand  how the electroweak scale could be separated from the scale of new physics without much  fine-tuning.  
This suggests that new physics should become manifest close to the TeV scale. 
Moreover, we expect that the new physics includes a symmetry protecting the higgs potential and that the electroweak breaking scale is  calculable in terms of microscopic parameters defining the fundamental theory. 
This kind of reasoning sets the main line of attack in physics beyond the standard model. 
 
One interesting possibility is that higgs is a pseudo-goldstone boson.
In such a scenario the higgs potential is protected by an approximate shift symmetry that arises after spontaneous breaking of a global symmetry.   
The pseudo-goldstone  mechanism has been already seen at work in high energy physics; it protects the masses of pions in QCD.  

In the pseudo-goldstone scenario, 
much of the low-energy features is fixed by  the pattern of the global symmetry and its spontaneous breaking.
However, several important issues, for example the scale of the global symmetry breaking, depends on the dynamics.
While it is  possible that this dynamics is weakly coupled and governed by another higgs sector, the more likely option is that it is strongly coupled (as in QCD). 

Some insight into strong dynamics can be gained by studying 5D theories in a warped background that are conjectured to be dual to strongly coupled theories \cite{adscftrs}. 
More precisely, a 4D pseudo-goldstone higgs can be realized in five dimensions as the so-called gauge-higgs unification scenario \cite{ho}. 
In 5D, the role of the higgs is played by the fifth component of a gauge field and the shift symmetry protecting the higgs potential originates from 5D gauge symmetry. 
The gauge symmetry, backed by 5D locality, is powerful enough to make the higgs potential fully calculable; divergent corrections cancel out to all orders it perturbation theory \cite{hoso}.
In consequence, the electroweak breaking scale can be unambiguously calculated in terms of the parameters of the 5D lagrangian. 
One can also address  further qualitative issues, such as corrections to the electroweak precision observables \cite{agco,caposa} or WW scattering amplitude in the resonance region \cite{faporo}.
Research along these line led to pinpointing classes of  models that pass all experimental tests \cite{agcopo,agcoda}, and the fine-tuning needed to achieve it can be quantified.

It should be stressed that all computations are performed on the 5D side, using the usual rules of (higher-dimensional) perturbation theory. 
In fact, gauge-higgs unification on its own is often viewed as a legitimate framework for physics beyond the standard model, independently of the holographic interpretation. Nevertheless,the 4D picture provides  more insight and a simple interpretation of the results obtained in 5D. 

Below  I concentrate on radiative generation of the higgs potential in gauge-higgs unification. 
I review the methods that make it possible to study this issue for a completely arbitrary warp factor.  
In fact, working with a general background is a simplification rather than a complication; in a specific background such as AdS one may easily drown in the sea of Bessel functions. 
This presentation is based on the results obtained in Refs. \cite{aa,mewa}.

\section{Methods}
\label{sec:1} 

Consider a 5D gauge theory propagating in a warped background with the line element given by  
\beq
ds^2 = a^2(y) dx_\mu dx^\mu - dy^2 . 
\eeq 
The fifth dimension is  an interval, $x_5 \in [0,L]$. 
We refer to $y = 0(L)$ as the UV(IR) brane. 
The function $a(y)$ is called the warp factor. 
We fix $a(0) = 1$. For $a(y) = 1$ we recover 5D flat space, while $a(y) = e^{- k y}$ corresponds to a slice of \ads. 
In the following discussion i keep the warp factor arbitrary.

The bulk contains gauge fields $A_M = A_M^\alpha T^\alpha$. 
The 5D gauge symmetry is broken to a subgroup by imposing Dirichlet boundary conditions: 
$A_\mu^a(0)  = 0$ and $A_\mu^b(L) = 0$ for some of the generators (while the remaining ones obey Neumann boundary conditions). 
Now, for a gauge field with Dirichlet boundary conditions on {\it both} branes, the fifth component contains a physical  mode that is massless at tree level.
This is our gauge-higgs field.  We assume that it acquires a vev:
\beq
\la A_5 \ra =   T^\ha {a^{-2}(y) \over \left (\int_0^L a^{-2} \right )^{1/2} } \ti v
\eeq
where we singled out one generator $T^\ha$  along which the vev resides. 
The normalization factors are chosen such that oscillations around the vev, $\ti v \to \ti v+ h(x)$, correspond to a canonically normalized scalar in the 4D effective theory. 
This field plays the role of the higgs boson.

Next,we consider a 5D fermion multiplet $\Psi$ charged under the gauge group. 
The quadratic action in the warped background reads  
\beq
S_5 = \int d^4 x \int_{0}^{L} dy  \left \{ 
 i a^3 \ov \Psi \gamma^\mu \pa_\mu \Psi 
 -  a^4 \ov \Psi (\gamma^5 \hat D_y  - M  ) \Psi 
  \right \} 
\eeq
where $\hat D_y = \hat \pa_y - i g_5 \la A_5^\ha \ra t^\ha$ and $\hat \pa_y  = \pa_y + 2 a'/a$. 
We expand the fermion in the basis of Kaluza-Klein (KK) mass eigenstates:  
\bea
\Psi_L(x,y) &=& f_{L,n}(\ti v,y) \Psi_{L,n}(x) 
\nn
\Psi_R(x,y)& =& f_{R,n}(\ti v,y) \Psi_{R,n}(x) 
\eea 
The profiles satisfy the equations of motion:
\bea
\left (\hat D_5 + M  \right  )f_{R,n} (\ti v,y)  &=& m_n(\ti v) a^{-1} f_{L,n} (\ti v,y)  
\nn
\left (-\hat D_5  + M  \right )f_{L,n}(\ti v,y)  &=& m_n(\ti v) a^{-1} f_{R,n} (\ti v,y) 
\eea 
that depend on the gauge-higgs vev. 
Furthermore, the profiles satisfy the boundary conditions that define the particular model.
The KK expansion allows us to rewrite the 5D quadratic action as a 4D action that is  diagonal in the KK basis: 
\beq 
S_5 = \int d^4 x   
 \ov \Psi_n(x) [i \gamma^\mu \pa_\mu  - m_n(\ti v) ]\Psi_n(x)  .
\eeq
The masses of the fermionic modes depend on the gauge-higgs vev.  
This implies that,  at the quantum level, the higgs will acquire a non-trivial potential. 
The potential is calculated from the usual Coleman-Weinberg formula: 
\beq
\label{cw5d}
V(\ti v) = - {N_c \over 4 \pi^{2}} \sum_n \int dp\, p^{3} \log \left (1 + {m_n^2(\ti v) \over p^2} \right ) 
\eeq 
where the summation goes over all fermionic eigenstates in the theory.
The 5D origin of this formula is manifest in the presence of the  infinite KK tower.   
Remarkably enough, one can rewrite Eq.\ref{cw5d} to a form where the KK summation is absent and where it resembles a 4D Coleman-Weinberg potential:    
\beq
V(\ti v) = - {N_c \over 4  \pi^{2}}  \int_0^\infty dp\, p^{3} \log \left ( \rho[-p^2]\right ) ,  
\eeq 
where the spectral function $\rho(s)$ is any analytic function whose simple zeros on the positive real axis encode the spectrum in the presence of the gauge-higgs vev: $\rho[m_n^2(\ti v)] = 0$. 
A tower of KK particles in 5D is a holographic manifestation of a composite structure in 4D.

In order to find the spectral function we need to solve the equations of motion with appropriate boundary conditions.
To this end,  we first introduce the auxiliary (hatted) profiles by 
\bea
\label{wr}
f_{L,n}(\ti v,y) &=& a^{-2}(y) e^{M y}  \Omega(y) \hat f_{L,n}(y)
\nn
f_{R,n} (\ti v,y) &=& a^{-2}(y) e^{-M y} \Omega(y) \hat f_{R,n}(y)
\eea 
The Wilson-line matrix is defined as   
\beq
\Omega(y) = e^{i g_5 \int_0^y \la A_5 \ra} 
\eeq
and its role is to rotate away the higgs vev from the equations of motion.
The hatted profiles satisfy simpler equations 
\bea
\pa_y \hat f_{R,n}  &=& m_n(\ti v) a^{-1} e^{2 M y} \hat f_{L,n}  
\nn
-\pa_y \hat f_{L,n} & =&  m_n(\ti v) a^{-1} e^{-2 M y} f_{R,n}  
\eea
that do not depend on $\ti v$ other than through the mass eigenvalues.
From the above follow the second order differential equations: 
\bea &
\left [ a e^{-2 M y} \pa_y ( a e^{2 M y} \pa_y ) + m^2 \right ] \hat f_{L,n} = 0
\nl
\left [ a e^{2 M y} \pa_y ( a e^{-2 M y} \pa_y ) + m^2 \right ] \hat f_{R,n} = 0
\eea  
with $m = m_n(\ti v)$
We denote two independent solutions of the first equation as $C_M(y)$ and $S_M(y)$ (consequently, the second is solved by $C_{-M}$ and $S_{-M}$). 
We pick up these solutions such that they satisfy  $C_M(0) = 1$, $C'_M(0) = 0$, $S_M(0) = 0$, $S'_M(0) = m$. 
The notation is to stress the similarity to  the familiar sines and cosines (to which these functions reduce for a flat warp factor and $M=0$). 
The warped generalization of $\sin' = \cos$ is 
$S_{M}'(y)   =   m a^{-1}(y) e^{-2 M y}  C_{-M}(y)$, 
$C_{M}'(y)   = - m a^{-1}(y) e^{-2 M y}  S_{-M}(y)$. 
The generalization of $\sin^2 + cos^2 = 1$ is the Wronskian $S_M(y) S_{-M}(y)+C_M(y) C_{-M}(y) = 1$.  
At small masses we can approximate $C_M = 1 - \co(m^2)$, $S_M = m \int_0^y a^{-1}(y') e^{- 2 M y'}$.

With this formalism at hand, the simple algorithm for computing spectral functions follows. 
1) Write down the auxiliary profiles: $\hat f_{L}^f = \alpha_C^f C_M -  \alpha_S^f S_M$, 
$\hat f_R = \alpha_C^f S_{-M} +  \alpha_S^f C_{-M}$, where the constant $\alpha$'s depend on the UV boundary conditions defined for a given flavour $\Psi^f$. 
2) Find the profiles $f(\ti v,y)$ from Eq. \ref{wr}. 
3) Write down IR boundary conditions and solve for $\alpha$'s. 
In the 5D models the solution exists for a discrete set of $m_n$.
The equation that sets the quantization condition can be employed as the spectral function.

\section{Applications: SO(5)}
\label{sec:2} 

We apply the general methods outlined in the previous section in the context of electroweak breaking driven by quarks with quantum numbers of the standard model top quark.
The gauge group is chosen to be $SO(5) \times U(1)_X$ as it is the simplest possibility that incorporates the electroweak group, the correct Weinberg angle and the custodial symmetry. 
The last is indispensable in 5D warped models in order to keep the Peskin-Takeuchi  T parameter under control.
$SO(5)$ has 10 generators: $T_L^a$ form the $SU(2)_L$ subgroup (identified with the standard model $SU(2)$), 
$T_R^a$ form the $SU(2)_R$ subgroup (identified with the custodial symmetry) and the remaining four generators $T_C^b$ belong to the $SO(5)/SO(4)$ coset. 
The gauge symmetry on the UV brane is reduced down to $SU(2)_L \times U(1)_Y$, the hypercharge being a combination of $T_R^3$ and $U(1)_X$.   
On the IR brane, the symmetry is reduced down to $SO(4) = SU(2)_L \times SU(2)_R$.
The four generators from the  $SO(5)/SO(4)$ coset have Dirichlet boundary conditions on both branes, so that the fifth components of the corresponding gauge fields host the higgs doublet. 
The vev is chosen  along the $T_C^4$ generator. 
The electroweak breaking scale is 
$v = f \sin (\ti v/f) $ where $ f^2 = 2/g_5^2 \int_0^L a^{-2}$ sets the global symmetry breaking scale. 

I will investigate the one-loop induced  higgs potential in the models with the top quark embedded in the spinorial $\bf 4$ representation of $SO(5)$ \cite{agcopo}. 
$\bf 4$ is the smallest $SO(5)$ representation and the spectral functions end up being less complicated than in models based on other $SO(5)$ representations.  
Although models based on the spinorial representation cannot be made fully realistic because of excessive contributions to the Zbb vertex \cite{agcoda}, they are most suitable for an illustrating purpose.  Spectral functions in the models with top quarks embedded in the fundamental $SO(5)$ representation are discussed in Ref. \cite{mewa}.

\subsection{Minimal model} 

Consider a 5D fermion field $Q$ with a bulk mass $M_Q$, transforming in the spinorial  $\bf 4$ representation of SO(5). 
The two SM top quark chiralities are embedded as 
\beq
Q =  ( q, \ q^c)   = ( t, \ b, \ t^c, \  b^c )   
\eeq 
The boundary conditions for the top quark fields\footnote{In the following discussion I ignore the bottom quarks whatsoever. In the model as it stands the bottom quarks must be either very heavy or degenerate in mass with the top quarks. Intuitively, once the bottom sector is properly realized (see Ref. \cite{agcopo}), we do not expect large contributions to the higgs potential because of  the small  bottom quark coupling to the higgs} are  
\beq
t_R(0) = t_L^c(0)  = 0 \quad t_R(L) = t_L^c(L)  = 0 . 
\eeq
In this simple set-up the we can readily display all step leading to the spectral function.
We first write down the auxiliary profiles that satisfy the UV boundary conditions: 
\beq
\hat f_{L,n}^t  =   \alpha_{t,n}  C_M(y) \quad
\hat f_{R,n}^t =  \alpha_{t,n}  S_{-M}(y)
\eeq 
\beq
\hat f_{L,n}^{t^c}  =   - \alpha_{t^c,n}  S_M(y) \quad
\hat f_{R,n}^{t^c} =  \alpha_{t^c,n}  C_{-M}(y)
\eeq 
To derive the full profiles $f(\ti v,y)$ we use Eq. \ref{wr} with the Wilson-line matrix for the spinorial representation given by: 
\beq
\Omega(L) = \left [\ba{cc}
\cos(\ti v/2f) & \sin (\ti v/2 f)
\\
\sin(\ti v/2f) & \cos (\ti v/2 f) \ea \right ]
\eeq 
The IR boundary conditions imply:
\bea 
\alpha_{t,n} \cos(\ti v/2f)   S_{-M}(L) + \sin (\ti v/2 f)\alpha_{t^c,n}  C_{-M}(L) &=& 0 
\nn
- \alpha_{t,n} \sin(\ti v/2f)   C_{M}(L) + \cos (\ti v/2 f)\alpha_{t^c,n}  S_{M}(L) &=& 0 
\nn
\eea 
The determinant of this equation is the quantization condition. 
We write the spectral function as   
\beq
\rho(m^2) =  1 -  {\sin^2(\ti v/2 f) \over  S_{M_Q}(L) S_{-M_Q}(L)}. 
\eeq 
As discussed in \cite{aa}, this spectral function can be  well approximated by 
\beq
\rho(-p^2) \approx  1 +  {y_t^2 f^2 \over \mkk^2 \sinh^2(p/\mkk)} \sin^2(\ti v/2 f)
\eeq 
where $y_t^2  = {L \int_0^L \, a^{-2}(y) \over 
\int_0^L a^{-1}(y) e^{-2M_Q y} \int_0^L a^{-1}(y) e^{2M_Q y}}$ plays the role of the top Yukawa coupling 
and $\mkk =  {1 \over \int_0^{L} a^{-1}(y)}$ sets the resonance scale. 
It follows that, at small momenta,  $\rho(-p^2) \approx   1 + y^2 f^2 \sin^2(\ti v/2 f)/p^2$, which is appropriate  for a 4D  top quark with the mass  
$m_t = y_t f \sin(\ti v/2 f)$. 
The effect of the resonance tower is to  exponentially suppress the higgs dependence of spectral function  for Euclidean momenta larger than $\mkk$. 
This implies that the coupling of the top quark to the higgs becomes very soft at high energies, which cuts off divergent corrections of the higgs potential. 
From the holographic point of view, the softness can be explained as a consequence of a composite structure of the higgs. 

Investigation of the higgs potential with this spectral function reveals a  minimum  at $\sin^2(\ti v/2 f) = 1$. 
This result can be simply understood. The higgs potential is dominated by the light top quark contribution whose quadratic divergence is cut off at the resonance scale, which leads to  $V(\ti v)\sim - m_t^2(\ti v) M_{KK}^2$.
Energetically the most favourable is to make the top quark as heavy as possible, which is achieved when $\sin^2(\ti v/2 f)$ is maximized.  
Thus the minimal model has too simple a structure to generate a realistic higgs potential 

\subsection{Shadow multiplet}

The problem with the higgs potential  of the minimal set-up can be solved by introducing the so-called shadow multiplet, that is another another 5D quark in a spinorial representation  
\beq
S =  (\ti q, \ \ti q^c) = ( \ti t, \ \ti b, \ \ti t^c, \  \ti b^c )   
\eeq 
with different boundary conditions in IR  
\beq
\ti t_R(0) = \ti t_L^c(0)  = 0 \quad  \ti t_L(L) = \ti t_R^c(L)  =0.
\eeq
At zero gauge-higgs vev the shadow multiplet does no  yield any light modes below the resonance scale  (hence the name ``shadow'', as opposed to ``light'').  
However, a massless modes appears for  $\sin^2(\ti v/2f) = 1$ as can be read off from the spectral function:  
\bea
\rho(m^2) =& 
\left (S_{M_Q}(L) S_{-M_Q}(L) -  \sin^2(\ti v/2f) \right ) 
\nn & 
\left (S_{M_S}(L) S_{-M_S}(L) -  \cos^2(\ti v/2f) \right )
\eea 
In this setup a correct electroweak breaking vacuum can be engineered. 
As before, the light multiplet tries to maximize  $\sin^2(\ti v/2f)$. 
On the other hand, the shadow multiplet prefers to maximize  $\cos^2(\ti v/2f)$.
Thus, the light and shadow compete and an equilibrium may be obtained for intermediate values of $\ti v/f$. 
The existence and location of the intermediate minimum depends on the bulk masses of the two multiplets (which determine the Yukawa couplings of the light top quarks to the higgs).  
For $M_Q = M_S$ the minimum falls in the middle,  for $\sin^2(\ti v/2f) = 1/2$. 
This is no good, as it corresponds to maximal electroweak breaking,
$v = f$, in which case the model is effectively higgsless (a light scalar field does exist but it does not unitarize WW scattering) with all the usual problems. 
However, with a certain degree of fine-tuning of the bulk masses  one can obtain a minimum at $\ti v/f \ll 1$, as  favoured by electroweak precision tests.
For example, a separation between $\ti v$ and $f$ is achieved for $y_{\ti t} \sim 1$ and $y_t^2 \sim 3/4$. 
To get $\ti v/f \sim 0.2$, the bulk masses have to be fine-tuned at the 10\% level.

\subsection{Mass mixing}

There is a simple way to complicate the previous model.  
The symmetries of the light-shadow set-up permit the mass terms on the IR brane,
\bea  
\cl = & - \delta(L) \ti M a^4(L) (\ov t_L, \ov b_L)\cdot (\ti t_R, \ti b_R)
\nn &   
 - \delta(L) \ti M_c a^4(L) (\ov t_R^c, \ov b_R^c) \cdot (\ti t_L^c, \ti b_L^c)
 + \hc,  
\eea 
that mix the two  multiplets.  
One reason to include the boundary masses is that they open more options for generating the correct electroweak breaking vacuum \cite{agcopo,mewa}.  

The spectral function takes the form 
\bea &  
\rho(m^2) =  
\nl  
[ C_{M_{S}}(L)  S_{-M_{Q}}(L)  +   |\ti M|^2 e^{2(M_Q-M_{S}) L}C_{M_{Q}}(L)  S_{-M_{S}}(L) ]
\nl
[C_{-M_S}(L)  S_{M_{Q}}(L)  +   |\ti M_c|^2 e^{2(M_{S}-M_Q) L} C_{-M_{Q}}(L)  S_{M_{S}}(L) ]
\nl 
- \sin^2(\ti v/2 f) [ \
|1 - \ti M \ti M_c^*|^2   
\nl
+ (1 - |\ti M \ti M_c |^2) (S_{M_{Q}}(L) S_{-M_{Q}}(L) -   S_{M_{S}}(L) S_{- M_{S}}(L))
 ] 
\nl
+ \sin^4(\ti v/2f)|1 - \ti M \ti M_c^*|^2  
\eea
In general, the model with the IR mass mixing becomes immune to analytic treatment and one should resort to numerical methods as in \cite{mewa}. This however goes beyond the scope of this presentation.

\vspace{1cm}

{\bf Acknowledgment:} I am partially supported by the European Community Contract MRTN-CT-2004-503369 for the years 2004--2008 
and by  the MEiN grant 1 P03B 099 29 for the years 2005--2007.

%
%

%

\begin{thebibliography}{999}
%
%


\bibitem{adscftrs}
  N.~Arkani-Hamed, M.~Porrati and L.~Randall,
  JHEP {\bf 0108}, (2001) 017
  [arXiv:hep-th/0012148].
  R.~Rattazzi and A.~Zaffaroni,
  JHEP {\bf 0104}, (2001) 021
  [arXiv:hep-th/0012248].
  M.~P\'erez-Victoria,
  JHEP {\bf 0105}, (2001) 064
  [arXiv:hep-th/0105048].

\bibitem{ho}
  Y.~Hosotani,
  Phys.\ Lett.\ B {\bf 126},(1983) 309 .
  Y.~Hosotani,
  Annals Phys.\  {\bf 190}, (1989) 233 .
  H.~Hatanaka, T.~Inami and C.~S.~Lim,
  Mod.\ Phys.\ Lett.\ A {\bf 13},(1998) 2601 
  [arXiv:hep-th/9805067].

\bibitem{hoso}
  Y.~Hosotani,
  arXiv:hep-ph/0607064.
  Y.~Hosotani, N.~Maru, K.~Takenaga and T.~Yamashita,
  arXiv:0709.2844 [hep-ph].

\bibitem{agco}
  K.~Agashe and R.~Contino,
  Nucl.\ Phys.\  B {\bf 742}, (2006) 59
  [arXiv:hep-ph/0510164].
\bibitem{caposa}
  M.~S.~Carena, E.~Ponton, J.~Santiago and C.~E.~M.~Wagner,
  Phys.\ Rev.\  D {\bf 76}, (2007) 035006
  [arXiv:hep-ph/0701055].

\bibitem{faporo}
  A.~Falkowski, S.~Pokorski and J.~P.~Roberts,
  arXiv:0705.4653 [hep-ph].

 \bibitem{agcopo}
  K.~Agashe, R.~Contino and A.~Pomarol,
  Nucl.\ Phys.\ B {\bf 719}, (2005) 165
  [arXiv:hep-ph/0412089].


\bibitem{agcoda}
  K.~Agashe, R.~Contino, L.~Da Rold and A.~Pomarol,
  Phys.\ Lett.\  B {\bf 641}, (2006) 62
  [arXiv:hep-ph/0605341].

\bibitem{aa}
  A.~Falkowski,
  Phys.\ Rev.\  D {\bf 75}, (2007) 025017
  [arXiv:hep-ph/0610336].
  A.~Delgado and A.~Falkowski, {\it unfinished}.
\bibitem{mewa}
  A.~D.~Medina, N.~R.~Shah and C.~E.~M.~Wagner,
  arXiv:0706.1281 [hep-ph].








\end{thebibliography}
%

\end{document}